\documentclass[aps,prb,reprint,showpacs,showkeys,groupedaddress]{revtex4-1}
\usepackage{amsmath}
\usepackage{graphicx}
\usepackage{rotating}
\DeclareMathOperator{\sech}{sech}
\usepackage{color}
\usepackage{caption}
\usepackage{subcaption}
\newcommand*{\rom}[1]{\uppercase\expandafter{\romannumeral #1}}
\begin{document}

%\preprint{}

\title{Probing non-locality of interactions in a Bose-Einstein Condensate using solitons}

\author{Abhijit Pendse}\thanks{abhijeet.pendse@students.iiserpune.ac.in}
%\thanks{}
\author{A. Bhattacharyay}\thanks{a.bhattacharyay@iiserpune.ac.in}
\affiliation{Indian Institute of Science Education and Research, Pune, Maharashtra, India 411008}

\date{\today}

\begin{abstract}
We consider a Bose-Einstein Condensate(BEC) with non-local inter-particle interactions. The local Gross-Pitaevskii(GP) equation is valid for the gas parameter $\nu=:a^{3}n_{0}<<1$, but for $\nu\rightarrow 1$, the BEC is described by modified GP equation(MGPE). We study the exact solutions of the MGPE describing bright and dark solitons. It turns out that the width of these non-local solitons has qualitatively similar behaviour as the modified healing length due to the non-local interactions of the MGPE. We also study the effect of the non-locality and gas parameter($\nu$) on the stability of the solitons using the Vakhitov –Kolokolov(VK) stability criterion. We show that these soliton solutions are indeed stable. Further, the stability of these soliton solutions gets enhanced due to the non-locality of interactions.
\end{abstract}

%\keywords{}

\maketitle

\section{Introduction}

Solitons are solitary, shape preserving travelling waveforms. These solutions are peculiar to non-linear differential equations and arise as a result of the balance between dispersion and non-linearity \cite{solitons_drazin}. Atomic Bose-Einstein Condensates(BECs) are described by the non-linear Gross-Pitaevskii(GP) equation which exhibits such soliton solutions \cite{soliton_tsuzuki}. The sign of the non-linearity in a BEC depends on the attractive or repulsive nature of the inter-particle interactions in a BEC. As a result, one obtains bright soliton solution in an attractive BEC and dark soliton solution in a repulsive BEC \cite{pethick,ps}. In a BEC, solitons have been realized in trapped atomic condensates \cite{solitons_burger, experiment_soliton_denschlag}. Solitons in BEC are of interest mainly due to the tunability by the control which atomic BEC systems provide in experiments. These control parameters include wide tunability of the s-wave scattering length in a BEC, the availablity of trapping techniques in order to modify the external potential and the achievement of multi-component BEC \cite{feshbach_inouye, becharmonictrap_fetter, twocomponent_soliton_wang, impurity_soliton_spielman}.

\par

The GP equation used to describe BECs considers s-wave scattering between bosons. The low energy scattering characteristics can be obtained by an integration over the actual interaction potential \cite{sakurai}. Hence, the actual interaction potential between bosons is replaced by a $\delta$-function pseudopotential to obtain the local GP equation. This shape-independent interaction approximation assumes the gas parameter $\nu=:a^{3}n_{0}<<1$, where $a$ is the s-wave scattering length and $n_{0}$ is the average density of the BEC. However, making use of the tunability of the s-wave interactions, there have been many experiments which achieve $\nu\sim 0.05$ or higher \cite{strongly_interacting_bec_papp, strongly_interacting_bec_claussen}. In this regime, the shape dependence of the s-wave interactions is expected to come into play. There have been various proposals to account for the shape dependence of the inter-boson interactions using a modified GP equation(MGPE) \cite{nonlocal_fu, nonlocal_pethick}. The MGPE is obtained by introducing an extra nonlinear term to the local GP equation which takes care of non-local interactions. The local GP equation admits a soliton solution, so the natural question would be to ask whether one can obtain a soliton solution to the MGPE as well and as to how it compares with the former\cite{sgarlata_jphysb}.

\par
An equation similar to the MGPE is used to describe a weakly non-local Kerr medium in optics \cite{nonlocal_krolikowski}. Although the form of the non-locality is similar, the origin of the non-locality is quite different in optics and BEC. Exact soliton solutions to the non-local Kerr medium have been obtained and studied, and are found to be stable \cite{pre_krolikowski}. However, to our knowledge, the implications of these solutions in a non-local BEC along with the stability considerations have not been studied. Also, in light of the stability quantifier proposed \cite{soliton_stability_sivan}, the effect of the non-local term to the stability of soliton solution in a BEC is of interest. In this paper we present a detailed analysis of the soliton solution of the MGPE in order to systematically manifest the role of microscopic interaction on such typical nonlinear states.

\par
The spectrum of elementary excitations for the local GP equation shows a phonon like behaviour $\omega\propto k$ for small wave-numbers($k$) and particle like behaviour $\omega\propto k^{2}$ for large $k$. The healing length is the length scale where there is a transition of this spectrum from phonon-like behaviour to particle-like behaviour. In the local GP equation the width of the soliton solution scales as $\xi_{0}=1/\sqrt{8\pi a n_{0}}$. For the MGPE, there is a change in the healing length and this modified healing length $\xi$ shrinks with increasing non-locality \cite{analoggravity_supratik} for a BEC with repulsive interactions. We show that the width of the dark soliton shows qualitatively similar behaviour as $\xi$.

\par
1-D solitons in local GP equation are known to be unstable to transverse small amplitude oscillations and as such they have to be strongly confined along the radial directions \cite{soliton_stability_muryshev}. Thus, soliton solutions are obtained only for effective 1-D BEC. The Vakhitov-Kolokolov(VK) stability condition ensures the stability of solitons for the 1-D local GP equation \cite{stability_vk}. This stability check is also of interest for the soliton solutions in the MGPE. The VK condition determines the stability of the solitons based on the sign of derivative of the momentum vs. speed graph for dark solitons and particle number vs. chemical potential graph for bright solitons. There has also been a proposal for a quantifier of soliton stability which depends on the magnitude of the aforementioned derivatives and not just the sign \cite{soliton_stability_sivan}. In this paper we show that according to these criterion, not only are the solitons in the MGPE stable, but their stability is enhanced by the non-locality parameter.

\par
We start by briefly describing the well known soliton solution in a local GP equation in Sec.\ref{sec:local_soliton}. Then, in Sec.\ref{sec:mgpe}, the MGPE is introduced. In Sec.\ref{sec:nonlocal_soliton_bright} we use the soliton solutions previously obtained by Krolikowski {\textit{et al.}} \cite{pre_krolikowski} and study the bright soliton solution for the case of non-local BEC. In Sec.\ref{sec:nonlocal_soliton_dark} the same is done for a dark soliton in a non-local BEC. We end with a discussion on the scope of the soliton solutions in a non-local BEC.
 
\section{Soliton solution in local GP equation}\label{sec:local_soliton}
Let us briefly review the well-known soliton solution in the local GP equation \cite{pethick, ps}. The GP equation with contact interactions($\nu<<1$) in the absence of an external potential in given by

\begin{equation}
\label{eq:contact_gp}
i\hbar\frac{\partial \psi(\bf{r},t)}{\partial t}=-\frac{\hbar^{2}}{2m}\nabla^{2}\psi(\bf{r},t)\pm g|\psi(\bf{r},t)|^{2}\psi(\bf{r},t),
\end{equation}

where the $+$ sign in the final term indicates a BEC with repulsive interactions and $-$ sign indicates one with attractive interactions. In the absence of the non-linear term, one would get a dispersive system such that any local disturbance in the system would disperse. The non-linearity counters this dispersion and leads to creation of localized structures like soliton \cite{solitons_drazin}. We want to look at 1-D soliton solutions of this equation. Therefore, we assume that $\psi({\bf{r}},t)$ varies only along 1 direction. Let that direction be the $z$-direction. This can be achieved in experiment, by strong harmonic confinement in the radial direction. Consider the form of the solution as $\psi({\bf{r}},t)=\sqrt{n_{0}}f(z,t)$, where $n_{0}$ is the uniform density of the BEC far away from the soliton center. The localized soliton solution has the boundary conditions $|f|\rightarrow0$, $df/dz \rightarrow 0$ as $z\rightarrow \pm \infty$ for a bright soliton. For dark soliton, the conditions are $|f|\rightarrow1$, $df/dz \rightarrow 0$ as $z\rightarrow \pm \infty$ . This means that for an attractive BEC, one gets a bright soliton solution, whereas a dark soliton solution is obtained for a repulsive BEC. The exact bright soliton solution is given by

\begin{equation*}
\psi(z,t)=\sqrt{n_{0}}\sech[(z-vt)/\sqrt{2}\xi_{0}] e^{\frac{it}{2\hbar}(\mu-4mv^{2})}e^{\frac{2imvz}{\hbar}},
\end{equation*}

 where $v$ is the soliton speed and $\xi_{0}=\hbar/\sqrt{2mgn}$ is the healing length. One can see that the bright soliton density far away from the soliton centre falls off to $0$. The bright soliton moves along the z-direction like a particle.

\par
For a BEC with repulsive inter-boson interactions, one can obtain an exact dark soliton solution given by

\begin{equation*}
\psi(z,t)=\sqrt{n_{0}}\Big[i\frac{v}{c}+\sqrt{1-\frac{v^{2}}{c^{2}}}\tanh\Big(\frac{z-vt}{\sqrt{2}\xi_{0}}\sqrt{1-\frac{v^{2}}{c^{2}}}\Big)\Big],
\end{equation*}

 where $c=\sqrt{gn_{0}/m}$ is the sound speed, $v$ is the soliton speed and $n_{0}$ is the uniform density far away from the soliton centre. Unlike the bright soliton where the density far from the soliton centre tends to zero, the density in a dark soliton heals to a uniform density far away from the soliton centre. Also, the dark soliton has a phase which depends both on the co-moving coordinate $(z-vt)/\xi_{0}$ and the speed $v$ of the soliton. Note that, soliton width in both cases scales as $\xi_{0}$.

%\par
%We next present the MGPE model and using the energy functional associated with the model, we give plausibility arguments on the expected change in soliton width as compared to the local GP equation.

\vspace{5mm}

\section{The MGPE}\label{sec:mgpe}
In this section, we consider a model to account for the non-locality of the s-wave interactions. This model considers an additional term to the local GP equation and is referred to as the MGPE in literature\cite{nonlocal_fu, nonlocal_pethick}. The MGPE, in the absence of an external potential, is given by

\begin{equation}
\label{eq:nonlocal_gp}
\begin{split}
i\hbar\frac{\partial \psi({\bf{r}},t)}{\partial t}=&-\frac{\hbar^{2}}{2m}\nabla^{2}\psi({\bf{r}},t)\pm g|\psi({\bf{r}},t)|^{2}\psi({\bf{r}},t)\\
&\pm g\:g_{2}\psi({\bf{r}},t)\nabla^{2}|\psi({\bf{r}},t)|^{2},
\end{split}
\end{equation}

where $g_{2}=\big(\frac{a^{2}}{3}-\frac{ar_{e}}{2}\big)$ and the $+$ sign implies a BEC with repulsive interactions and $-$ sign implies a BEC with attractive interactions. The effective range of the interactions is captured by $r_{e}$. This equation is valid only for $g_{2}>0$. This means that $r_{e}<2a/3$, since $r_{e}=2a/3$ would correspond to scattering by a hard sphere\cite{nonlocal_pethick}. As stated in the introduction, the non-local correction would usually become important at a higher value of the gas parameter $\nu$. The energy functional of the MGPE is given by\cite{energy_pendse}

\begin{widetext}
\begin{equation}
\label{eq:energy_mgpe}
F= \int \vec{dr}[ \frac{\hbar^{2}}{2m}|\nabla\psi({\bf{r}},t)|^{2} \pm \frac{g}{2}|\psi({\bf{r}},t)|^{4} \pm \frac{gg_{2}}{2}|\psi({\bf{r}},t)|^{2}\nabla^{2}|\psi({\bf{r}},t)|^{2}].
\end{equation}
\end{widetext}

 As done for the local GP equation, we look at 1-D soliton solutions for the MGPE in what follows.

\section{Bright soliton in an attractive MGPE}\label{sec:nonlocal_soliton_bright}
A bright soliton solution can be obtained in an attractive BEC. This corresponds to a self-focussing nonlinearity with non-locality for an optical medium, for which exact implicit solutions were obtained by Krolikowski {\textit{et al.}}\cite{pre_krolikowski}. The solutions obtained shall be used to see the behaviour of bright solitons in a BEC in what follows. The 1D GP equation for attractive BECs can be written as

\begin{equation}
\label{eq:nonlocal_gp_attractive}
\begin{split}
i\hbar\frac{\partial \psi(z,t)}{\partial t}=&-\frac{\hbar^{2}}{2m}\frac{\partial^{2}\psi(z,t)}{\partial z^{2}}-g|\psi(z,t)|^{2}\psi(z,t)\\
&-g\:g_{2}\psi(z,t)\frac{\partial^{2}}{\partial z^{2}}|\psi(z,t)|^{2},
\end{split}
\end{equation}

where $g$ is taken to be positive and the attractive nature of BEC is expressed by putting a minus sign in the equation itself. Notice that since $g$ is positive, so is $a$ in the above equation. This is to say that $a$ physically is negative, but the negative sign is taken out of $a$ and written it explicitly in the equation.

\begin{figure}[h]
	\includegraphics[width=0.4\textwidth]{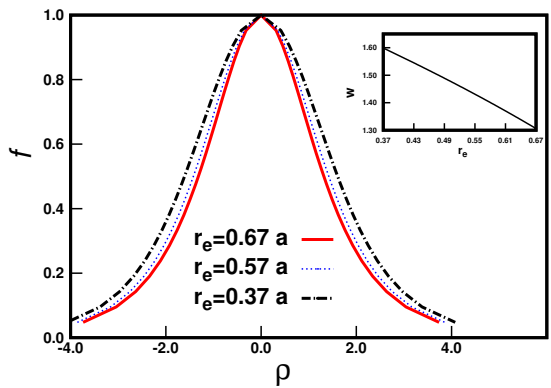}
 
\caption{The figure shows the density profile of the bright soliton for different values of $r_{e}$ and $\alpha=1$. The inset shows the variation of the width of bright soliton with variation in $r_{e}$(in units of $a$) for $\nu=0.1$.}
\label{fig:density_profile_bright}
\end{figure}

\par 

Let us consider here a system with finite size which is very large as compared to the soliton width and the healing length. Let $N$ be the number of bosons in the system and $V$ be its volume. For the state with uniform density, the density for the major part of the volume would be $n_{0}=N/V$, barring the healing at the boundaries. This healing is neglected in the analysis that follows, since the system size is taken to be large as compared to the healing length. To obtain a bright soliton solution, moving with a speed $v$, the $z$ coordinate is scaled as $\rho=(z-vt)/(\xi_{0}\sqrt{2})$, where $\xi_{0}=\hbar/\sqrt{2mgn_{0}}$. Further, we write $\psi(z,t)=\sqrt{n_{0}}f(\rho)e^{\frac{it}{2\hbar}(\alpha gn_{0}-4mv^{2})}e^{\frac{i2mvz}{\hbar}}$ where $n_{0}$ is the uniform density in the absence of soliton and $\alpha$ is the factor which accounts for the presence of soliton in the system. This turns the above equation to

\begin{equation}
\label{eq:nonlocal_gp_attractive_scaled}
\frac{\partial^{2}f}{\partial\rho^{2}}+2(|f|^{2}-\frac{\alpha}{2})f+\frac{g_{2}}{\xi_{0}^{2}}f\frac{\partial^{2}|f|^{2}}{\partial\rho^{2}}=0.
\end{equation}

This equation can be integrated once, to give

\begin{equation}
\Big(\frac{df}{d\rho}\Big)^{2}=\frac{(\alpha-f^{2})f^{2}}{1+2\big(\frac{g_{2}}{\xi_{0}^{2}}\big)f^{2}} \; .
\label{eq:bright_derivative}
\end{equation}

The central peak density of the bright soliton can be evaluated from the above equation. This can be done by setting $df/d\rho=0$ for $\rho=0$. This gives the peak density as $f^{2}(0)=\alpha$. The factor $\alpha$ is evaluated by evaluating the energy functional ($F$) and then from $dF/dN$. This means that the central density of the soliton would depend on number of particles ($N$) and volume ($V$) of the system.
\par
 A further integral of the above equation leads to the implicit soliton solution \cite{pre_krolikowski}

\begin{equation}
\label{eq:nonlocal_bright_sol}
\pm\rho= \tanh^{-1}{(\sigma/\sqrt{\alpha})}+\sqrt{\gamma} \tan^{-1}{(\sigma\sqrt{\gamma})},
\end{equation}

where, $\gamma=2g_{2}/\xi_{0}^{2}$ and $\sigma^{2}=\frac{\alpha-f^{2}}{1+\gamma f^{2}}$. In this implicit solution, there appear square roots of $\alpha$ and $\gamma$ which may have $+$ or $-$ sign. However, for the solution to exist, it is essential that only the $+$ or the $-$ square root of $\gamma$ are considered consistently. For the root of $\alpha$, only the $+$ square root can be taken for the solution to exist, otherwise the sign preceding the $\tanh^{-1}$ term would become negative. Since there are two instances when $\gamma$ appears on the RHS, taking different sign for two $'\gamma'$s would imply a change of sign between the $\tanh^{-1}$ and $\tan^{-1}$ terms. This implicit expression would no longer be a solution of Eq.(\ref{eq:nonlocal_gp_attractive_scaled}) as can be verified by differentiating it twice.

\par
By plugging different values of $f$ from $0$ to the peak density $\alpha$ in the RHS of Eq.(\ref{eq:nonlocal_bright_sol}), one obtains Fig.(\ref{fig:density_profile_bright}) which shows the soliton density as a function of $\rho$. Here, $\alpha$ is set equal to $1$ for the sake of simplicity. The value of $r_{e}$ is bounded above by $r_{e}\sim0.67a$ as mentioned in the introduction. From Fig.(\ref{fig:density_profile_bright}) and its inset, one can see the variation in width of the soliton with $r_{e}$.

\par

Using the bright soliton solution in Eq.(\ref{eq:nonlocal_bright_sol}), the energy for a bright soliton on top of the uniform density state can be evaluated using the expression for energy functional in Eq.(\ref{eq:energy_mgpe}). Here too, $\alpha=1$ for simplicity. Since the integrand involves second derivative of $|f|^{2}$, it is convenient to change the integration variable from $\rho\rightarrow f$. This can be done using Eq.(\ref{eq:bright_derivative}) giving $d\rho=df\frac{\sqrt{1+\gamma f^{2}}}{f\sqrt{1-f^{2}}}$. Also, using Eq.(\ref{eq:bright_derivative}), one can find $d^{2}|f|^{2}/d\rho^{2}=2\times [(df/d\rho)^{2}+f(d^{2}f/d\rho^{2})]$. Further, subtracting the energy of the uniform density state from this integral gives the soliton energy. This gives the expression for energy integral as

\begin{widetext}
\begin{equation*}
\label{eq:energy_bright_sol}
%\begin{split}
E=\frac{gn^{2}}{2}-\sqrt{2}\xi_{0}gn^{2}\int_{0}^{1}\Big\{\Big[\frac{f^{2}(1-f^{2})}{1+\gamma f^{2}}+(2f^{2}-1-f^{4})
-\frac{\gamma}{2}\Big(\frac{2f^{4}-3f^{6}-\gamma f^{8}}{1+\gamma f^{2}}\Big)+8U^{2}f^{2}\Big]\times\frac{\sqrt{1+\gamma f^{2}}}{f\sqrt{1-f^{2}}}\Big\}df. 
%\end{split}
\end{equation*}
\end{widetext}

\begin{figure}[h]
	\includegraphics[width=0.4\textwidth]{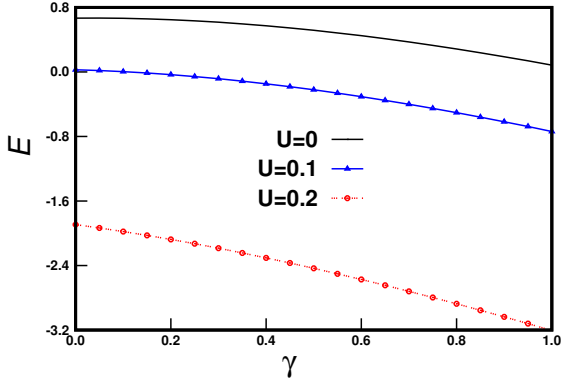}
 
\caption{The figure shows the variation of the energy over the ground state uniform density of the bright soliton with change in $\gamma$, where $\gamma=2g_{2}/\xi_{0}^{2}$ for some values of $U$ and $\alpha=1$.}
\label{fig:energy_bright}
\end{figure}

Fig.(\ref{fig:energy_bright}) shows the variation of energy with respect to $\gamma$ for certain values of $U=v/(c\sqrt{2})$. This figure shows that the effect of the non-locality in an attractive BEC is to decrease the magnitude of energy of the soliton as compared to the uniform density state for a fixed speed.

\par
\begin{widetext}

\begin{figure}[h]
\captionsetup[subfigure]{oneside,margin={-1.0cm,-1.0cm}}

   \begin{subfigure}[b]{0.4\textwidth}
     \rotatebox{0}{	\includegraphics[width=\textwidth]{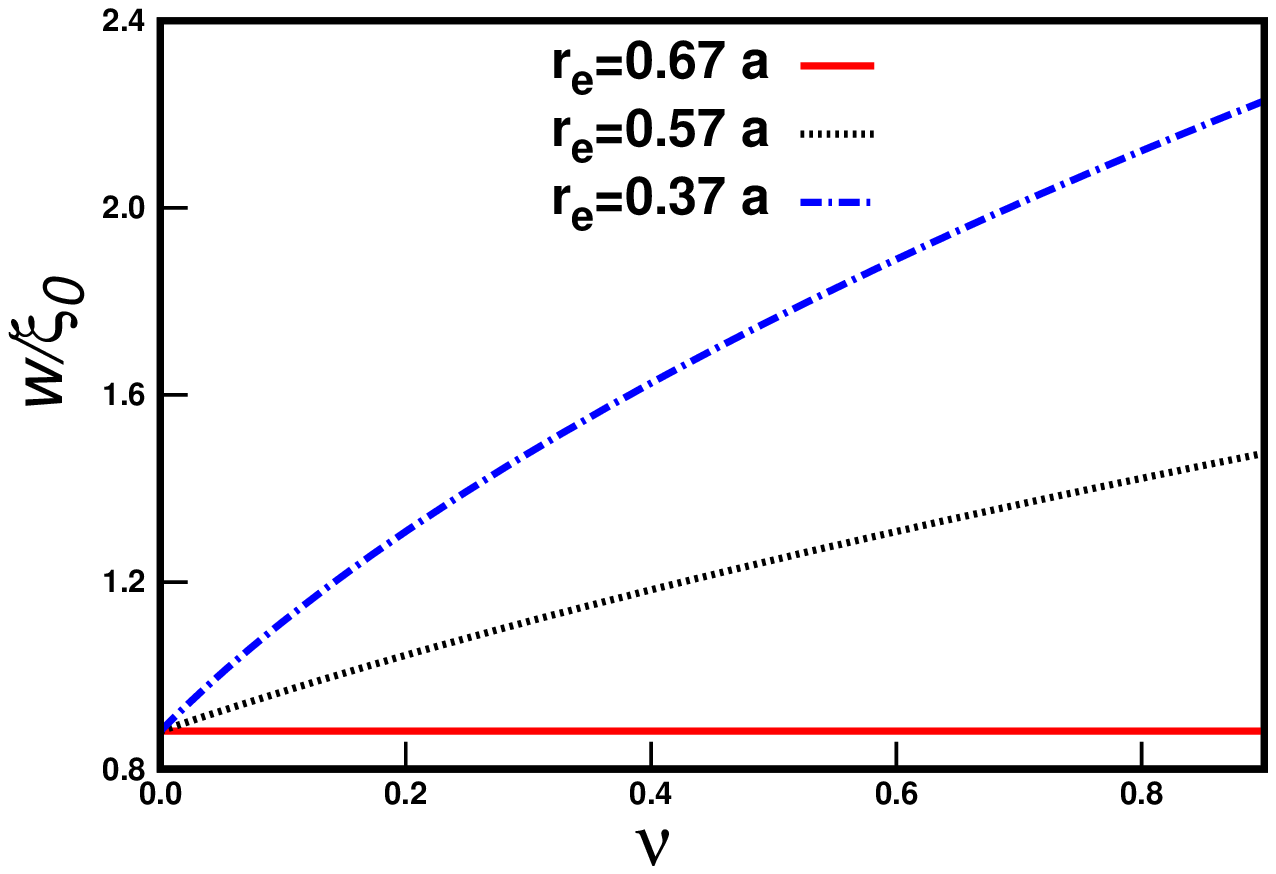}}
  \caption{U=0, graph scaled by $\xi_{0}$}
    \end{subfigure}
\hspace{1cm}
   \begin{subfigure}[b]{0.4\textwidth}
     \rotatebox{0}{	\includegraphics[width=\textwidth]{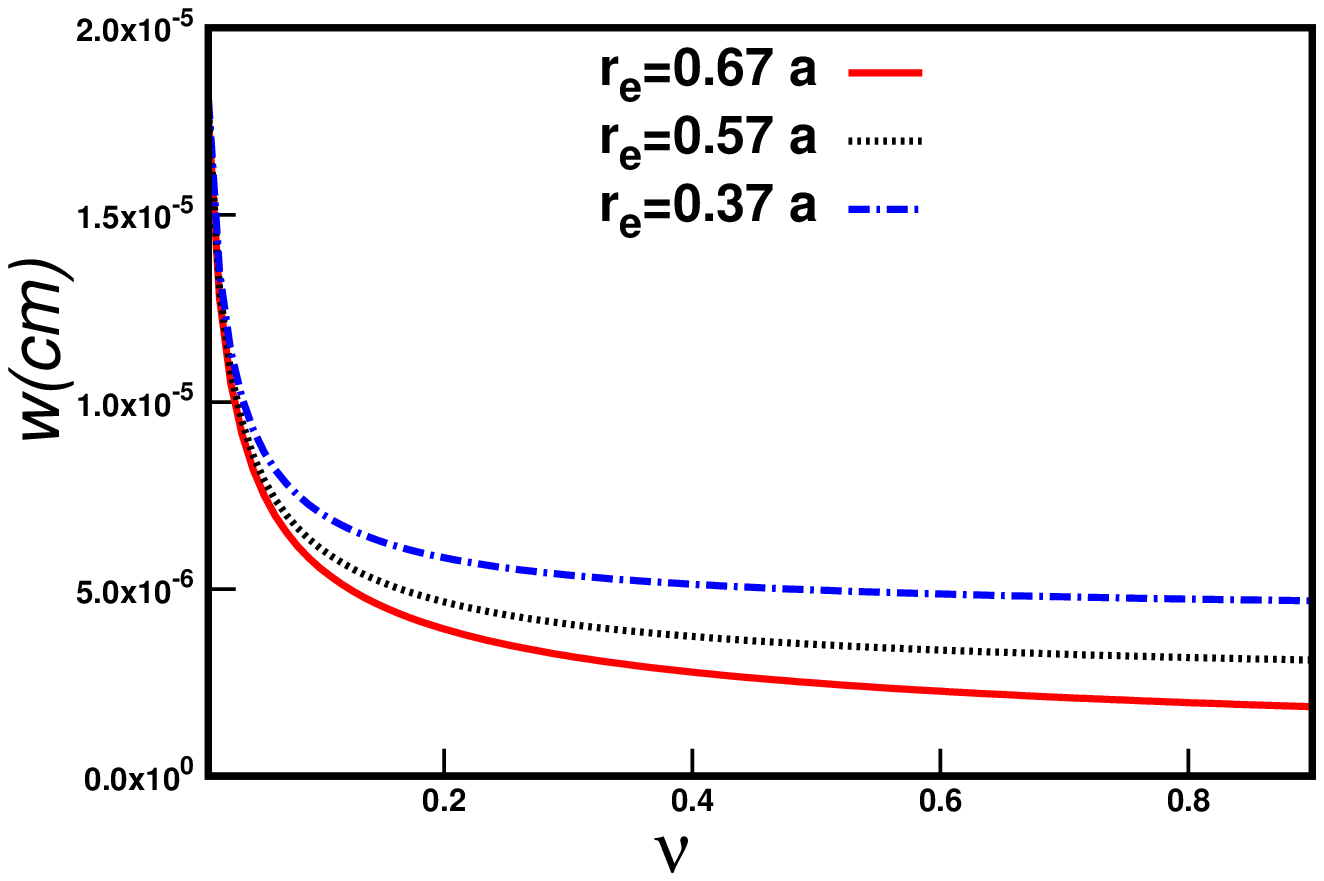}}
  \caption{U=0, graph unscaled}
    \end{subfigure}

\caption{Figures show the variation of the width of the bright soliton with change in the gas parameter $\nu$ with $\alpha=1$. In subfigure $(a)$, the width is scaled with the healing length $\xi_{0}$, whereas in subfigure $(b)$, the width is not scaled.}
\label{fig:bright_soliton_width}
\end{figure}

\end{widetext}

We define the width of the bright soliton to be the $\rho$ where the soliton density falls to half of its peak density. Substituting $f^{2}=1/2$ in Eq.(\ref{eq:nonlocal_bright_sol}), we take $r_{e}=\beta a$ giving $\gamma=2g_{2}/\xi_{0}^{2}=16\pi \nu(\frac{1}{3}-\frac{\beta}{2})$. Here, $\beta$ is a constant proportionality factor. Thus, one obtains, in Fig.(\ref{fig:bright_soliton_width}), a plot of the width of the bright soliton as a function of $\nu$ for different values of $r_{e}$. Note that, the width is independent of the speed $v$ of the soliton. Also note the non-linear behaviour for $g_{2}\neq 0$ of the scaled width as $\nu$ is increased.

\par
The width can be analytically found by putting $f^{2}=1/2$ in Eq.(\ref{eq:nonlocal_bright_sol}). This gives us the width to be 

\begin{equation*}
\frac{w}{\xi_{0}}=\tanh^{-1}\Big(\sqrt{\frac{2\alpha-1}{\alpha(2+\gamma)}}\Big)+\sqrt{\gamma}\tan^{-1}\Big(\sqrt{\frac{\gamma(2\alpha-1)}{2+\gamma}}\Big).
\end{equation*}

 Here, $w/\xi_{0}$ is written so as $\rho$ contains a scaling factor of $\xi_{0}$. In the absence of the non-local correction, $\gamma=0$ and $w/\xi_{0}=\tanh^{-1}(\sqrt{\frac{2\alpha-1}{2\alpha}})$, meaning $w/\xi_{0}$ is a constant. This implies that the length scale of width of the soliton is $\xi_{0}$. However, in the presence of the non-local correction, the width gets modified and the modified length scale is given by $B\xi_{0}$, where $B=\tanh^{-1}\big(\sqrt{\frac{2\alpha-1}{\alpha(2+\gamma)}}\big)+\sqrt{\gamma}\tan^{-1}\big(\sqrt{\frac{\gamma(2\alpha-1)}{2+\gamma}}\big)$.

\vspace{1cm}

\par

By considering small amplitude oscillatory disturbance around the uniform density ground state of the form, $\psi({\bf{r}},t)=(\sqrt{n_{0}}+ue^{i{\bf{k.r}}}e^{-i\omega t}+ve^{-i{\bf{k.r}}}e^{i\omega t})e^{-i\mu t/\hbar}$, one can obtain the dispersion relation for small amplitude oscillations as $\omega=\pm k\sqrt{-\frac{gn}{m}+\frac{\hbar^{2}k^{2}}{4m^{2}}}$ for the local GP equation. The value of $k$ for which the dispersion relation turns from phonon-like($\omega\sim k$) to particle like($\omega\sim k^{2}$) is $\sim \xi_{0}$. Since an attractive BEC is unstable, the dispersion relation becomes imaginary for small $k$. Nevertheless, one can define the healing length to be the length scale when magnitude of the phonon-like ($\omega\sim k$) term becomes comparable to the magnitude of the particle-like term ($\omega\sim k^{2}$). A similar analysis can be done for the MGPE to obtain the spectrum of elementary excitations which comes out to be $\omega=\pm k\sqrt{-\frac{gn}{m}+\big(\frac{\hbar^{2}}{4m^{2}}+\frac{\hbar^{2}gg_{2}n}{m}\big)k^{2}}$\cite{analoggravity_supratik}. The length scale of phonon-particle transition of elementary excitation spectrum can be found by equating the magnitude of terms which go as $k$ and the term which goes as $k^{2}$. This length scale turns out to be $\xi=\xi_{0}\sqrt{1+\gamma}/\sqrt{2}$. This change in the healing length indicates that there has to be a broadening of the width of the bright soliton in the presence of non-local interactions and, here, such a change can be seen. So, in the presence of non-local interactions the width of the bright soliton somewhat follows the modified healing length. In other words, the widening of the width of a bright soliton from the healing length is a proof of existence of non-local interactions and corresponding change in the healing length.

\par
Having seen the behaviour of soliton width and energy in the presence of the non-local correction, the next natural step is to check the stability of the bright soliton solution. This stability condition is given by the sign of $dN/d\mu$ \cite{bright_stability_ladtke}, where $N=\int{|\psi(z,t)|^{2}}$ is the total number of particles in the BEC and $\mu=\alpha gn_{0}/2$. The bright soliton is stable if $dN/d\mu>0$. The bright soliton solution above can be used to obtain

\begin{equation*}
N=\hbar\sqrt{\frac{n_{0}}{mg}}\Big(\sqrt{\alpha}+\frac{1+\gamma\alpha}{\sqrt{\gamma}}\tan^{-1}(\sqrt{\gamma\alpha})\Big).
\end{equation*}

\vspace{1cm}

This gives

\begin{equation}
\label{eq:stability_slope_bright}
dN/d\mu\propto\frac{1}{\sqrt{\alpha}}+\sqrt{\gamma}\tan^{-1}(\sqrt{\gamma\alpha}).
\end{equation}

\par

\begin{figure}[h]
	\includegraphics[width=0.4\textwidth]{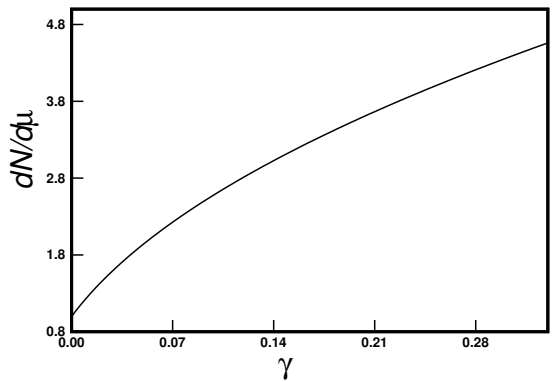}
 
\caption{The figure shows the variation of $dN/d\mu$ as we change $r_{e}$(in scale of $a$). Here, $\alpha=1$ and $\nu=0.1$.}
\label{fig:stability_bright}
\end{figure}

As explained before, $\sqrt{\gamma}$ and $\sqrt{\alpha}$ are positive and hence $\tan^{-1}(\sqrt{\gamma\alpha})$ is also positive, which ensures that $dN/d\mu>0$. This shows that bright solitons are stable. Fig.(\ref{fig:stability_bright}) shows the variation $dN/d\mu$ as a function of $\gamma$. This graph shows that the bright soliton solution is indeed stable. There has been a proposal of quantifying the stability based on $|dN/d\mu|$ \cite{soliton_stability_sivan}. It says that the soliton stability increases with increase in the magnitude of $|dN/d\mu|$. In this regard, it can be seen from Fig.(\ref{fig:stability_bright}) that the soliton stability is enhanced by increasing the non-locality of interactions.

\section{Dark soliton in a repulsive MGPE}\label{sec:nonlocal_soliton_dark}

Let us now look at the dark soliton solutions which can be obtained in a repulsive BEC. This corresponds to a self-defocussing non-linearity with non-locality in an optical medium. The 1D non-local GP equation for a repulsive BEC is given as

\begin{equation}
\label{eq:nonlocal_gp_repulsive}
\begin{split}
i\hbar\frac{\partial \psi(z,t)}{\partial t}=&-\frac{\hbar^{2}}{2m}\frac{\partial^{2}\psi(z,t)}{\partial z^{2}}+g|\psi(z,t)|^{2}\psi(z,t)\\
&+g\:g_{2}\psi(z,t)\frac{\partial^{2}}{\partial z^{2}}|\psi(z,t)|^{2},
\end{split}
\end{equation}

As for the attractive BEC case in the previous section, let's start by considering $N$ bosons occupying a volume $V$. The uniform density state has $n_{0}=N/V$. The healing at the boundaries may be neglected by assuming that the system is large compared to the healing length. To find moving solutions, it is convenient to consider a moving frame of reference $s=(z-vt)/\sqrt{2}\xi_{0}$ and write $\psi=\psi(s)e^{-i\mu t/\hbar}$. The form of $\psi(s)$ is taken as $\psi(s)=\sqrt{n_{0}}\sqrt{\lambda(s)}e^{i\phi(s)}$. Using such a form of $\psi$, Eq.(\ref{eq:nonlocal_gp_repulsive}) gives two equations for real and imaginary parts as

\begin{equation}
\begin{split}
&2\lambda(1-\gamma\lambda)\frac{d^{2}\lambda}{ds^{2}}-\Big(\frac{d\lambda}{ds}\Big)^{2}-8\lambda^{2}(\lambda-\alpha)\\
&-4\lambda^{2}\Big(\frac{d\phi}{ds}\Big)^{2}+8U\lambda^{2}\frac{d\phi}{ds}=0;\\
&\\
&\frac{d}{ds}\Big[\lambda\Big(\frac{d\phi}{ds}-U\Big)\Big]=0,
\end{split}
\label{eq:nonlocal_gp_repulsive_2}
\end{equation}

where $\gamma=2g_{2}/\xi_{0}^{2}$ and $U=\sqrt{2}mv\xi_{0}/\hbar$. Since, dark soliton solutions are localized with an intensity minima at $s=0$, we integrate the equations above. Putting the boundary conditions of $d\lambda/ds=0$ for $s=0$ and $s\rightarrow \pm \infty$, one gets from the above equation the central dip $\lambda(0)=U^{2}$ and the background density $\lambda(s\rightarrow\pm\infty)=\alpha$. The lower of the two equations gives, $d\phi/ds=U$ when integrated. This can be used to integrate the upper equation giving

\begin{equation*}
\Big(\frac{d\lambda}{ds}\Big)^{2}=\frac{4(\lambda-U^{2})(\alpha-\lambda)^{2}}{1-\gamma\lambda}.
\end{equation*}

The above equation can then be integrated to give the following implicit solution for a dark soliton \cite{pre_krolikowski}

\begin{equation}
\label{eq:nonlocal_dark_sol}
\begin{split}
\pm s=&\frac{1}{\delta_{0}}\tanh^{-1}{\frac{\delta}{\delta_{0}}}+\sqrt{\gamma}\tan^{-1}{(\delta\sqrt{\gamma})}\\
\\
\\
\pm\phi(s)=&\tan^{-1}{\frac{\delta}{U}}-U\sqrt{\gamma}\tan^{-1}{(\delta\sqrt{\gamma})}.
\end{split}
\end{equation}

where $\delta^{2}=\frac{\lambda(s)-U^{2}}{1-\gamma\lambda(s)}$ and $\delta_{0}^{2}=\frac{\alpha-U^{2}}{1-\gamma}$. Here, $n_{0}$ is the background density over which one observes the dark soliton solution. As for the bright soliton solution, there appears square root of $\gamma$ which may have $+$ or $-$ sign. As before, only take the $+$ or the $-$ square root of $\gamma$ can be taken consistently. Taking different signs for two $'\gamma'$s would imply a change of sign between the $\tanh^{-1}$ and $\tan^{-1}$ terms which would no longer be a solution of Eq.(\ref{eq:nonlocal_gp_repulsive_2}).

\par
\begin{figure}[h]
	\includegraphics[width=0.4\textwidth]{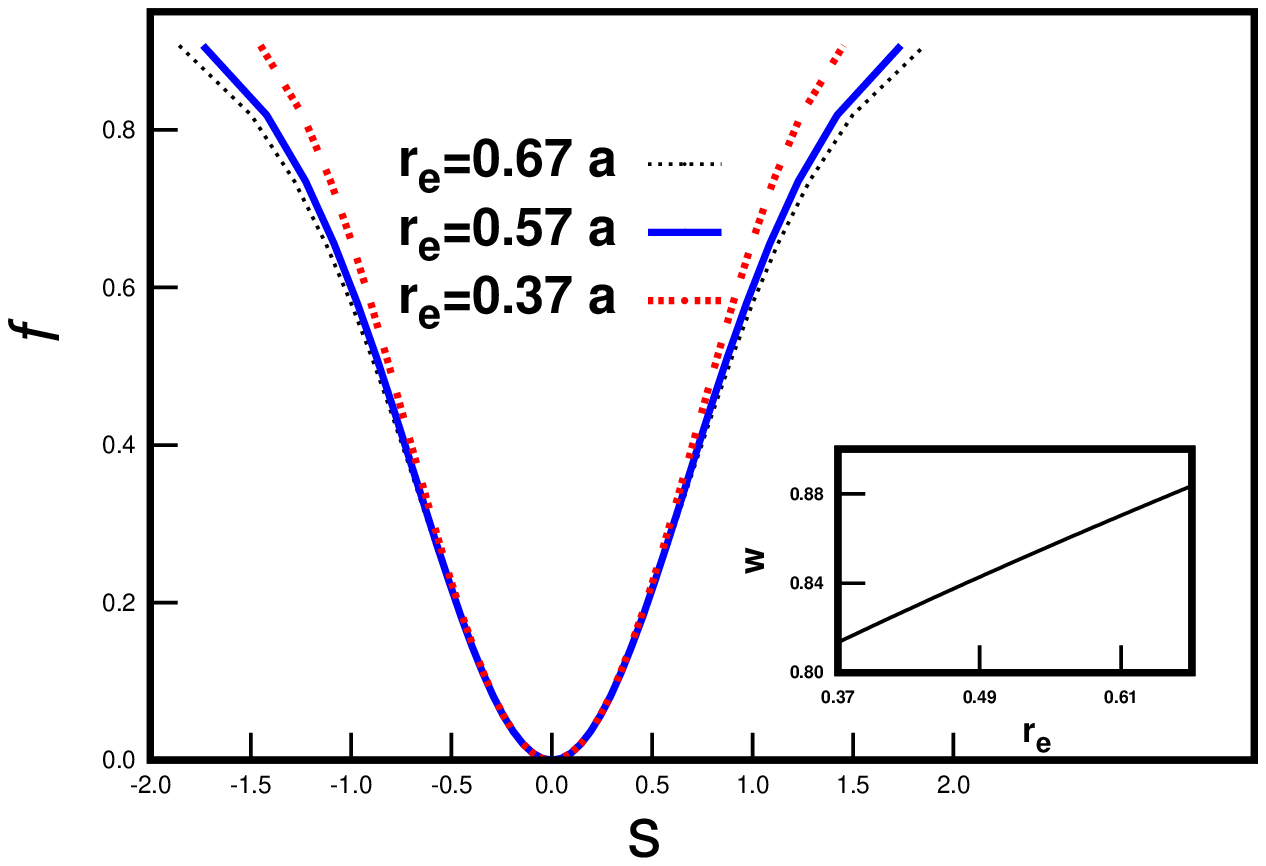}
 
\caption{The figure shows the density profile of the dark soliton for different values of $r_{e}$ and $\alpha=1$ for $U=0$. The inset shows the variation of the width of dark soliton with variation in $r_{e}$(in units of $a$) for $\nu=0.1$ for $U=0$.}
\label{fig:density_profile_dark}
\end{figure}

 Fig.(\ref{fig:density_profile_dark}) shows the density profile for a dark soliton with $U=0$ for different values of $r_{e}$, with the inset showing variation of the width with $r_{e}$ for $U=0$. This profile shows that non-locality changes the width of the dark soliton. However, while the width of bright soliton decreases with increasing $r_{e}$, the width of the dark soliton increases with an increase in $r_{e}$.

\par
As for the bright soliton, the energy for a dark soliton can be evaluated using the expression for energy functional in Eq.(\ref{eq:energy_mgpe}). Change of coordinates as for the bright soliton is employed. The coordinate change in this case is $s\rightarrow \lambda$ where $ds=d\lambda \;(\sqrt{1-\gamma\lambda})/[2(\alpha-\lambda)\sqrt{\lambda-U^{2}}]$. This gives, by setting $\alpha=1$ as before, the energy integral as

\begin{widetext}
\begin{equation*}
\begin{split}
E=\sqrt{2}\xi_{0}gn_{0}^{2}\int_{U^{2}}^{1}&\Big\{\frac{1}{2\lambda}\Big[\frac{(1-\lambda)\sqrt{\lambda-U^{2}}}{\sqrt{1-\gamma\lambda}}\Big]+\frac{\lambda U^{2}\sqrt{1-\gamma\lambda}}{2(1-\lambda)\sqrt{\lambda-U^{2}})}+\frac{(1-\lambda)\sqrt{1-\gamma\lambda}}{2\sqrt{\lambda-U^{2}}}\\
&+\frac{\gamma\lambda}{2}\Big(\frac{1-U^{2}(\gamma-2)-\lambda(3+U^{2}\gamma)+2\gamma\lambda^{2}}{\sqrt{\lambda-U^{2}}(1-\gamma\lambda)^{3/2}}\Big)\Big\}d\lambda
\end{split}
\end{equation*}
\end{widetext}

\vspace{1cm}

\begin{figure}
	\includegraphics[width=0.4\textwidth]{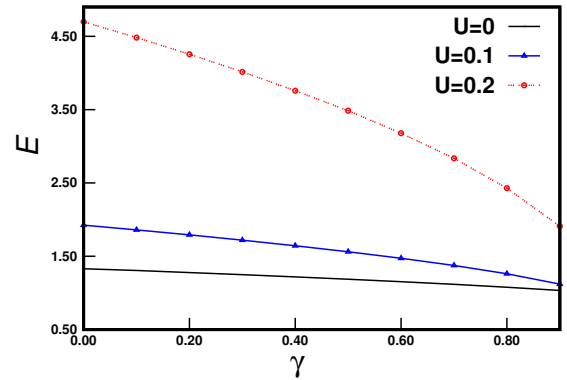}
 
\caption{The figure shows the variation of the energy over the ground state uniform density of the dark soliton with change in $\gamma$, where $\gamma=2g_{2}/\xi_{0}^{2}$ for some values of $U$ and $\alpha=1$.}
\label{fig:energy_dark}
\end{figure}

 Fig.(\ref{fig:energy_dark})shows the variation of energy with respect to $\gamma$ for certain values of $U$. From this figure, it can be seen that the effect of the non-locality is to decrease the magnitude of energy of the soliton above the uniform density state of a repulsive BEC.

\par

\begin{widetext}

\begin{figure}[h]
\captionsetup[subfigure]{oneside,margin={-1.0cm,-1.0cm}}

   \begin{subfigure}[b]{0.4\textwidth}
     \rotatebox{0}{	\includegraphics[width=\textwidth]{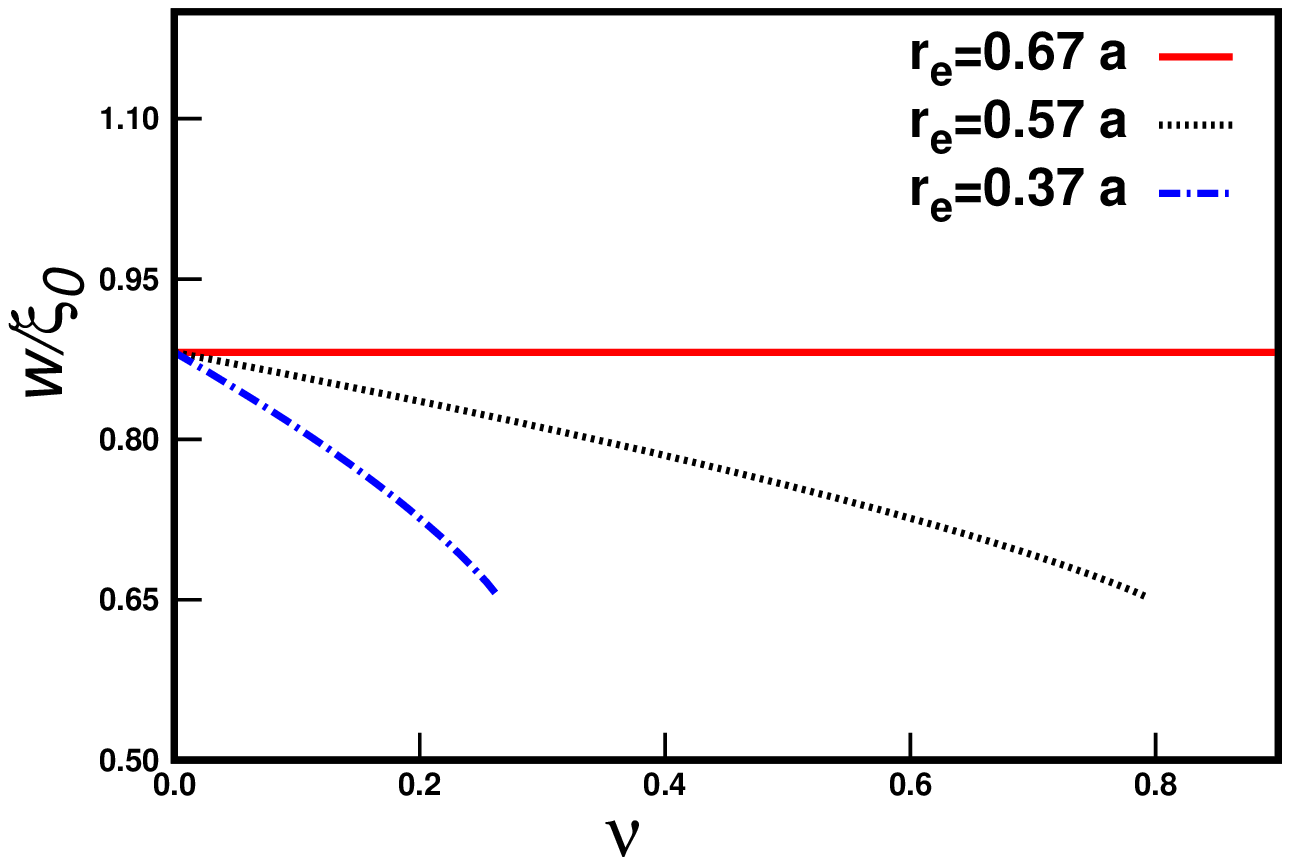}}
  \caption{U=0, graph scaled by $\xi_{0}$}
    \end{subfigure}
\hspace{1cm}
   \begin{subfigure}[b]{0.4\textwidth}
     \rotatebox{0}{	\includegraphics[width=\textwidth]{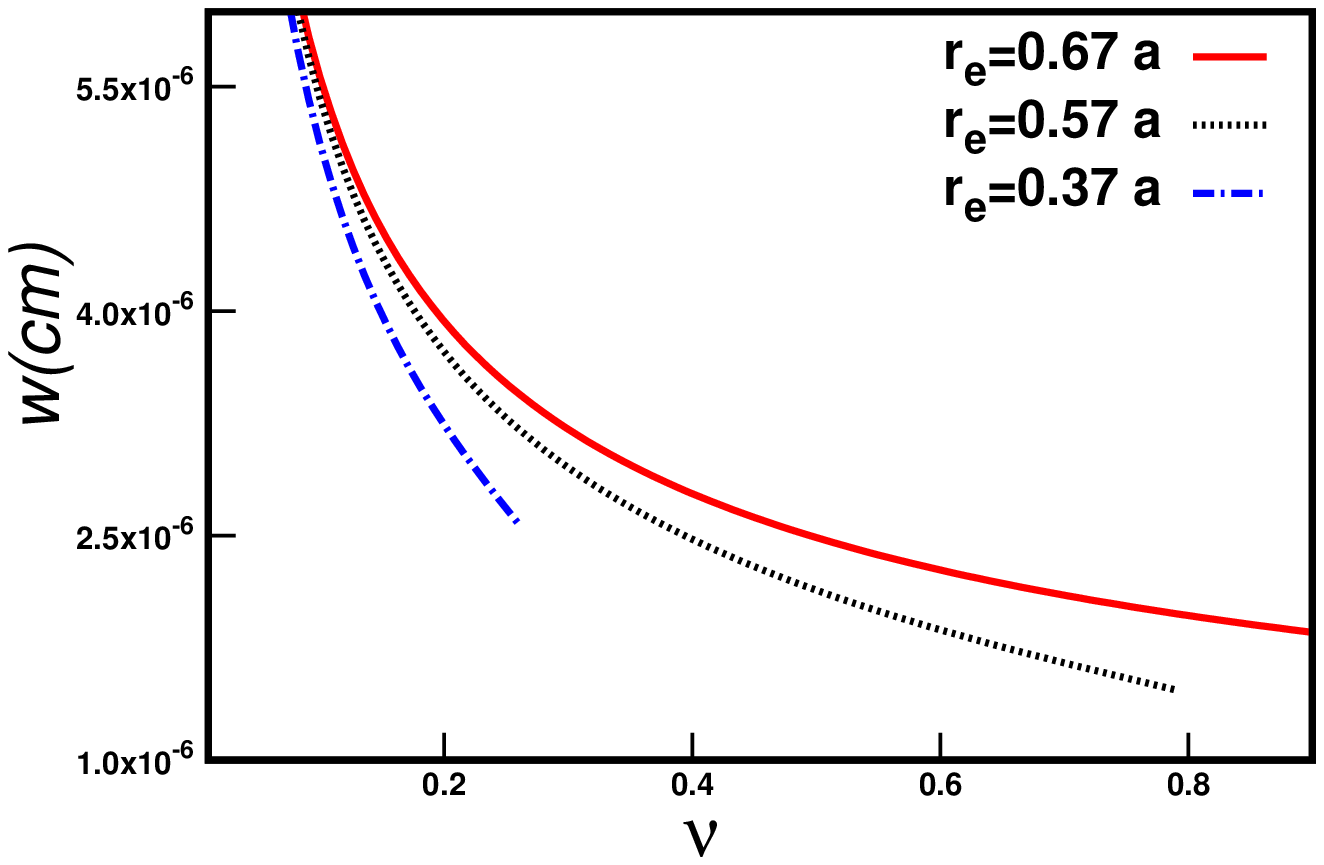}}
  \caption{U=0, graph unscaled}
    \end{subfigure}

\caption{Figures show the variation of the width of the black soliton with change in the gas parameter $\nu$ for $\alpha=1$. In subfigure $(a)$, the width is scaled with the healing length $\xi_{0}$, whereas in subfigure $(b)$, the width is not scaled.}
\label{fig:dark_soliton_width}
\end{figure}

\end{widetext}

Fig.(\ref{fig:dark_soliton_width}), shows the effect of non-locality on the width of the dark soliton for $U=0$. The incomplete graphs for $r_{e}<0.67a$ beyond certain values of $\nu$ are due to the fact that $\delta_{0}^{2}=\frac{1}{1-\gamma}$ becomes negative and hence $\delta_{0}$ becomes imaginary. This indicates that the implicit dark soliton solutions given by Eq.(\ref{eq:nonlocal_dark_sol}) exist only till a certain value of $\nu$ for a given value of $r_{e}$.

\vspace{1cm}

\par
 The change in width for a stationary dark soliton($U=0$) can be looked at analytically as follows. For the sake of simplicity, let us consider $r_{e}\propto a$, such that $g_{2}=ca^{2}$. Since $g_{2}=\frac{a^{2}}{3}-\frac{ar_{e}}{2}$, $0\leq c\leq 1/3$ because $g_{2}>0$ for a repulsive BEC. Further, since  $\xi_{0}=1/\sqrt{8\pi a n_{0}}$, we can write $\gamma$ as, $\gamma=2g_{2}/\xi_{0}^{2}=16\pi c\nu$. With $U=0$, $\delta^{2}=\frac{\lambda}{1-\gamma\lambda}$ and $\delta_{0}^{2}=\frac{1}{1-\gamma}$. Let us define the width to the point where the soliton density is half of the density at the point far off from the central dip. That s, the width of the soliton is the value of $s$(say $w/\xi_{0}$) when $\lambda=1/2$. This gives the width of the black soliton as

\begin{equation*}
\frac{w}{\xi_{0}}=\sqrt{\frac{1-\gamma}{\alpha}}\:\tanh^{-1}\Big(\sqrt{\frac{1-\gamma}{\alpha(2-\gamma)}}\Big)+\sqrt{\gamma}\:\tan^{-1}\Big(\sqrt{\frac{\gamma}{2-\gamma}}\Big),
\end{equation*} 

 Note that $s=w/\xi_{0}$ is taken because $s$ involves scaling by $\xi_{0}$. Had it been the local GP equation i.e., $\gamma=0$, one would have obtained $w=\tanh^{-1}\Big(\sqrt{\frac{1}{2}}\Big)$. This implies that $w/\xi_{0}$ is a constant, meaning that $\xi_{0}$ is the length scale for solitons, in the case of local GP dynamics. However, when $\gamma\neq 0$, it can be seen that the scale is different from $\xi_{0}$ and depends on $\gamma$. For a finite $\gamma$, $w/(A\times\xi_{0})$ is a constant, where $A=\sqrt{1-\gamma}\:\tanh^{-1}\Big(\sqrt{\frac{1-\gamma}{2-\gamma}}\Big)+\sqrt{\gamma}\:\tan^{-1}\Big(\sqrt{\frac{\gamma}{2-\gamma}}\Big)$. This implies that the new length scale is $A\xi_{0}$.

\par
As before doing a small amplitude oscillation analysis around the ground state of the form $\psi({\bf{r}},t)=(\sqrt{n_{0}}+ue^{i{\bf{k.r}}}e^{-i\omega t}+ve^{-i{\bf{k.r}}}e^{i\omega t})e^{-i\mu t/\hbar}$, one can obtain the dispersion relation for small amplitude oscillations as $\omega=\pm k\sqrt{\frac{gn}{m}+\frac{\hbar^{2}k^{2}}{4m^{2}}}$ for the local GP equation. Note here that since the repulsive is stable, $\omega$ is real. The healing length as before is $\sim \xi_{0}$. Doing such an analysis for the MGPE with repulsive interactions gives the modified healing length as $\xi=\xi_{0}\sqrt{1-\gamma}/\sqrt{2}$. This change in the healing length indicates that there has to be a shrinking in the width of the dark soliton in the presence of non-local interactions which can be observed here, as for the bright soliton. Hence, for dark solitons too, the width of the dark soliton somewhat follows the modified healing length in the presence of non-local interactions.

\par
However, the width of the soliton in the presence of non-local interactions although qualitatively follows the modified healing length $\xi$, the width is not exactly scaled by $\xi$. It scales with $c\xi$, where $c$ is a $\gamma$-dependent function.

\par
\begin{figure}[h]
	\includegraphics[width=0.4\textwidth]{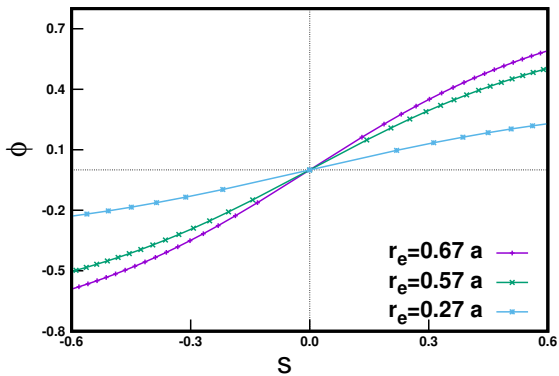}
 
\caption{The figure shows the variation of phase of the dark soliton $\phi(s)$ as a function of $s$ for a few values of $r_{e}$. The gas parameter is fixed at $\nu=0.1$, $\alpha=1$ and $U=0.8$.}
\label{fig:dark_soliton_phase}
\end{figure}

One of the popular methods of obtaining dark solitons experimentally is the phase imprinting method \cite{experiment_soliton_denschlag, phase_imprinting_biao}. From Eq.(\ref{eq:nonlocal_dark_sol}), it is evident that along with the spatial density variation, dark solitons are accompanied by a spatial variation of phase of the wave function, given by $\phi(s)$. The phase imprinting method, as the name suggests, uses a laser which is off-resonant with regards to the excitations in an atomic BEC and imprints the required phase for the generation of dark soliton. For a stationary soliton($U=0$), it can seen from Eq.(\ref{eq:nonlocal_dark_sol}) that there is a sudden phase jump at $s=0$ from $-\pi/2$ for $s<0$ to $\pi/2$ for $s>0$. It is phase jumps like these which have to be imprinted. For $U=0$, the factor $\gamma$ doesn't play a role. However, as $U$ increases, the variation of the phase with respect to $s$ is significantly altered by $\gamma$ and hence the phase to be imprinted for soliton generation changes with $\gamma$. Fig.(\ref{fig:dark_soliton_phase}) shows the phase variation of the dark soliton as a function of $s$ for few different values of $r_{e}$, with $U=0.8$.

% Expanding this modified healing length($\xi$) for small $\gamma$, we get

%\begin{equation*}
%\xi=\frac{\xi_{0}}{\sqrt{2}}[1-\frac{\gamma}{2}-\frac{\gamma^{2}}{8}- ...].
%\end{equation*}

%We had defined the width of the soliton to be the value of $s$ where the density becomes $1/2$ of the background density, i.e. $\lambda=1/2$. Let us define the width to be the value of $s$ where $\lambda=h$, where $h$ is an arbitrary number with the restriction $0<h<1$. Using this definition, the width is given by $\widetilde{w}=\xi_{0}B$, where $B=\sqrt{1-\gamma}\tanh^{-1}\big(\sqrt{\frac{h(1-\gamma)}{1-(h\gamma)}}\big)+\sqrt{\gamma}\tan^{-1}\big(\sqrt{\frac{h\gamma}{1-h\gamma}}\big)$. The expression for $\widetilde{w}$ can be expanded in power series for small $\gamma$ as

%\begin{equation*}
%\begin{split}
%\widetilde{w}=\xi_{0}&\Big[\tanh^{-1}(\sqrt{h})+\frac{\gamma}{2}\big(\sqrt{h}-\tanh^{-1}(\sqrt{h})\big)\\
%&+\frac{\gamma^{2}}{24}\big(3\sqrt{h}+h^{\frac{3}{2}}-3\tanh^{-1}(\sqrt{h})\big)+....\Big].
%\end{split}
%\end{equation*}

%\vspace{5cm}

%For the local GP equation, the same length scale $\xi_{0}$ shows up in both the small amplitude oscillations and the soliton width. However, we can see that for the MGPE this is not so in general. For the MGPE, the behaviour of $\xi$ and $\widetilde{w}$ becomes similar in order $\gamma^{1}$ when $h\sim0.924$. However, for a general $h$, the two length scales given by $\xi$ and $\widetilde{w}$ differ. Thus the correction to the local GP equation distinguishes the length scale of small amplitude oscillations and large amplitude solitons.

\par
As for the bright soliton, the stability of the dark soliton solution of the MGPE is of interest. The stability of dark solitons is provided by the Vakhitov-Kolokolov(VK) conditions which determine the stability based on the sign of $dQ/dU$, where $Q$ is the normalized momentum of the soliton given by
\begin{equation*}
\begin{split}
Q=\frac{i}{2}\int_{-\infty}^{\infty}[\psi(z,t)&\partial_{z}\psi^{*}(z,t)-\psi^{*}(z,t)\partial_{z}\psi(z,t)]\\
&\:\times\Big[1-\frac{n_{0}}{|\psi(z,t)|^{2}}\Big] dz
\end{split}
\end{equation*}

\par

Dark solitons are stable if $dQ/dU>0$ according to the VK condition. Further, as for the bright soliton, there is a proposal for a stability quantifier using the magnitue of $dQ/dU$. It states that the stability is enhanced as the value of $dQ/dU$ increases beyond $0$ \cite{soliton_stability_sivan}.

\par

Using the dark soliton solution, the expression for $Q$ is given by

\begin{equation*}
\begin{split}
Q=&-2\alpha\tan^{-1}\Big(\frac{\delta_{0}}{U}\Big)+(\alpha-U^{2})\frac{U}{\delta_{0}}\\
&+\frac{U}{\sqrt{\gamma}}\{1+\gamma(2\alpha-U^{2})\}\tan^{-1}(\delta_{0}\sqrt{\gamma}).
\end{split}
\end{equation*}

The integration is done by using the change of coordinates from $z\rightarrow \lambda$ as before.

Using this expression for $Q$, one can evaluate $dQ/dU$, which is given by

\begin{equation}
\begin{split}
\frac{dQ}{dU}&=\frac{2U^{4}\gamma-2U^{2}\alpha-\alpha U^{2}\gamma+3\alpha^{2}}{\delta_{0}(\alpha-U^{2}\gamma)}\\
&+\frac{U^{2}(1-U^{2}\gamma+2\alpha\gamma)}{\delta_{0}[-1+(1+U^{2}-\alpha)\gamma]}\\
&+\Big(\frac{1-3U^{2}\gamma+2\alpha\gamma}{\sqrt{\gamma}}\Big)\tan^{-1}{(\delta_{0}\sqrt{\gamma})}.
\end{split}
\label{eq:dark_stability}
\end{equation}

To bear in mind is the fact that $\delta_{0}$ is a function of U.

\begin{figure}[h]
	\includegraphics[width=0.4\textwidth]{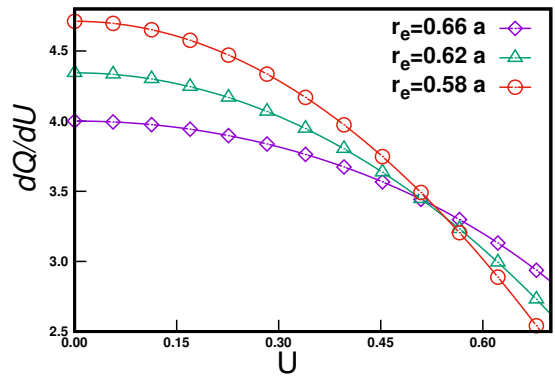}
 
\caption{The figure shows the value of $dQ/dU$ as a function of $U$ for a few values of $r_{e}$ . The gas parameter is fixed at $\nu=0.5$ and $\alpha=1$.}
\label{fig:stability_dark}
\end{figure}

Fig.(\ref{fig:stability_dark}) shows the variation of $dQ/dU$ with respect to $U$ for different values of the effective range $r_{e}$. These graphs are plotted for $\nu=0.5$ and $\alpha=1$. This graph shows that the solitons are stable even for such large values of $\nu$. Note that even for higher and lower value of $\nu$, one can use Eq.(\ref{eq:dark_stability}) to draw plots similar to Fig.(\ref{fig:stability_dark}) to show that solitons are stable for a range of $\nu$ values. However, at larger values of $\nu$, the higher order corrections would come into picture. For lower values of $\nu$, only the lower order correction used in MGPE is sufficient and as such the solitons are stable using the VK conditions.

\par
Fig.(\ref{fig:stability_dark}) shows that the VK stability condition is satisfied for  all $U$ for $\alpha=1$. A simple check can be done to show that $dQ/dU>0$ for all $\alpha$ for $U=0$. This can be done by putting $U=0$ in Eq.(\ref{eq:dark_stability}), which gives $dQ/dU=(3\alpha/\delta_{0})+[(1+2\alpha\sqrt{\gamma}\tan^{-1}(\delta_{0}\sqrt{\gamma}))/\sqrt{\gamma}]$, which is $>0$.

\vspace{5mm}

\par
 As for the bright solitons, $|dQ/dU|$ increases for a fixed value of gas parameter as we increase $\gamma$( i.e. decrease $r_{e}$) for $U=0$. For higher values of $U$, the value of the slope shows a more complicated behaviour with the graphs crossing each other. However, from Fig.(\ref{fig:stability_dark}) one can say with certainty that the increased nonlocality provides additional stability to the dark soliton \cite{soliton_stability_sivan} for $U$ close to $U=0$.

\section{Discussion}

In this paper, we have studied the 1-D soliton solutions for a modified Gross-Pitaevskii equation(MGPE). The MGPE takes into account the non-locality of s-wave interactions in a Bose-Einstein Condensate(BEC). Using the exact solutions obtained by Krolikowski {\textit{et al.}}, we have studied the behaviour of these solutions for a BEC. As the soliton solution is obtained due to a balance between dispersion and non-linearity, the introduction of correction term to the GP equation changes the length scale of the solitons. Consequently, the width of the solitons change as well. This change in the length scale of soliton width shows a similar behaviour to the modified healing length for the MGPE. The change in the width can be experimentally verified as a confirmation of the existence of non-locality of interactions in a BEC. An important consequence of it is that the value of effective range of interactions $r_{e}$ can be experimentally determined.

\par
The soliton solutions of the MGPE are stable even as one increases the gas parameter($\nu$) to small but finite values. The non-locality of interactions in a BEC imparts additional stability to the solitons, considering the proposal of quantifying soliton stability \cite{soliton_stability_sivan}. Given this, it would be interesting to see the effect of transverse modes which make the soliton unstable, on the soliton solutions of the MGPE. The non-locality of interactions may impart extra stability in the transverse direction as well, allowing for weaker transverse trapping as compared to the local GP case. We shall investigate this feature in a future study.

\vspace{1cm}

\begin{acknowledgements}
AP would like to thank the Council of Scientific and Industrial Research(CSIR), India for the funding provided.
\end{acknowledgements}

\bibliography{references_soliton.bib}

\end{document}